%% file: shear_bands_23-1-2015arxiv.tex
\documentclass[11pt]{article}

\usepackage[top=1.0in,right=1.0in,left=1.0in,bottom=1.75in]{geometry}
\usepackage{amssymb,amsbsy}
\usepackage{amsmath}
\usepackage{amsthm}
\usepackage{graphics,graphicx}
\usepackage[T1]{fontenc}
\usepackage{color}
\usepackage{subfigure}
\usepackage[affil-it,auth-sc]{authblk}
\usepackage{stmaryrd}
\usepackage{booktabs}
\usepackage{array}

\newcolumntype{L}[1]{>{\raggedright\let\newline\\\arraybackslash\hspace{0pt}}m{#1}}
\newcolumntype{C}[1]{>{\centering\let\newline\\\arraybackslash\hspace{0pt}}m{#1}}
\newcolumntype{R}[1]{>{\raggedleft\let\newline\\\arraybackslash\hspace{0pt}}m{#1}}

\input{definitions}

\newcommand{\ljump}{\llbracket}
\newcommand{\rjump}{\rrbracket}
\newcommand{\jump}[1]{\ljump #1 \rjump}

\graphicspath{{./}{./figures/}}

\title{Strain localization and shear banding in ductile materials}

\date{}

\author[]{N. Bordignon, A. Piccolroaz, F. Dal Corso and D. Bigoni}

\affil[]{DICAM, University of Trento, Italy}

\begin{document}

\maketitle

\begin{abstract}
\noindent
A model of a shear band as a zero-thickness nonlinear interface is proposed and tested using finite element simulations.
An imperfection approach  is used in this model where a shear band, that is assumed to lie in a ductile matrix material (obeying von Mises plasticity with linear hardening), is present from the beginning of loading and is considered to be a zone in which yielding occurs before the rest of the matrix.
This approach is contrasted with a perturbative approach, developed for a J$_2$-deformation theory material, in which the shear band is modelled to emerge at a certain stage of a uniform deformation.
Both approaches concur in showing that the shear bands (differently from cracks) propagate rectilinearly under shear loading and that a strong stress concentration should be expected to be present at the
tip of the shear band, two key features in the understanding of failure mechanisms of ductile materials.
\end{abstract}

{\it Keywords: }

\section{Introduction}

\begin{figure}[p]
\centering
\includegraphics[width=80mm]{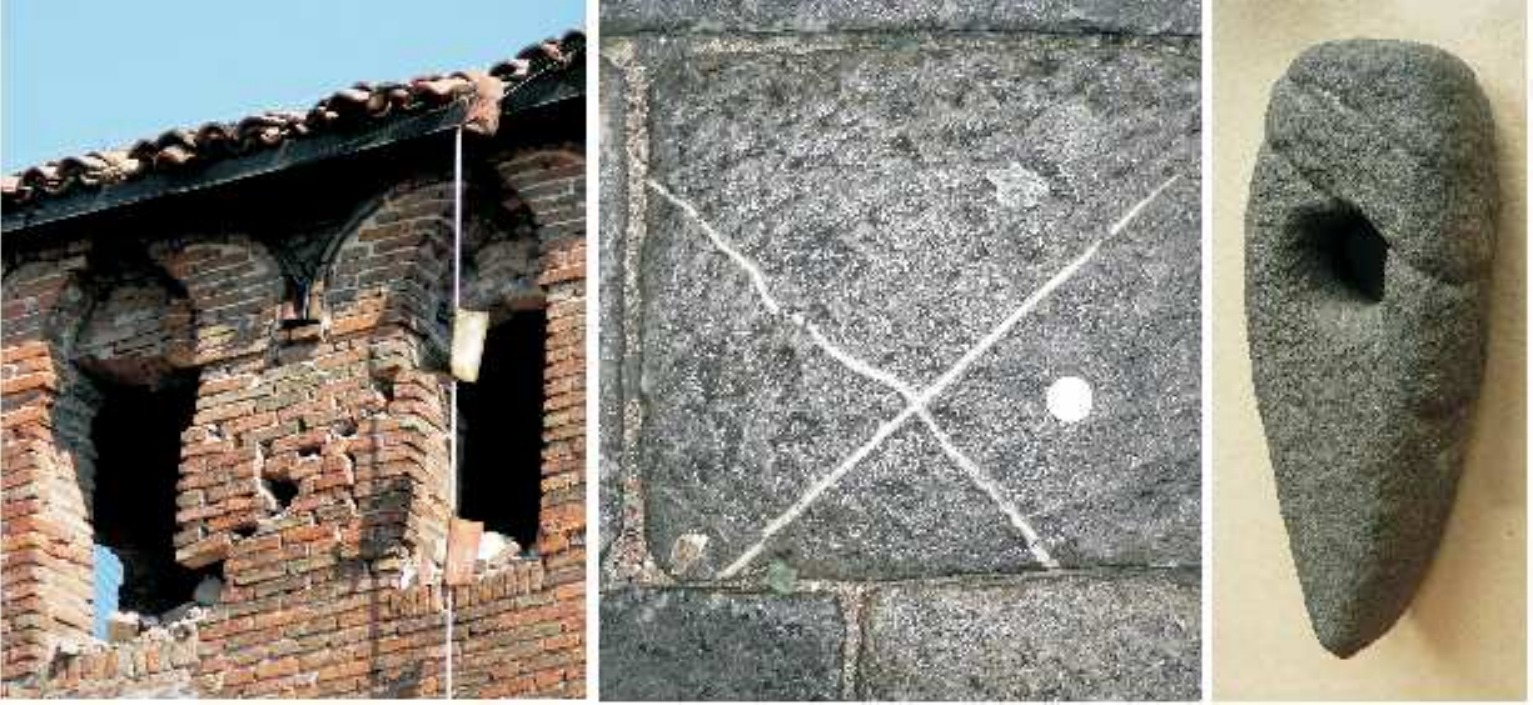} \\
\includegraphics[width=80mm]{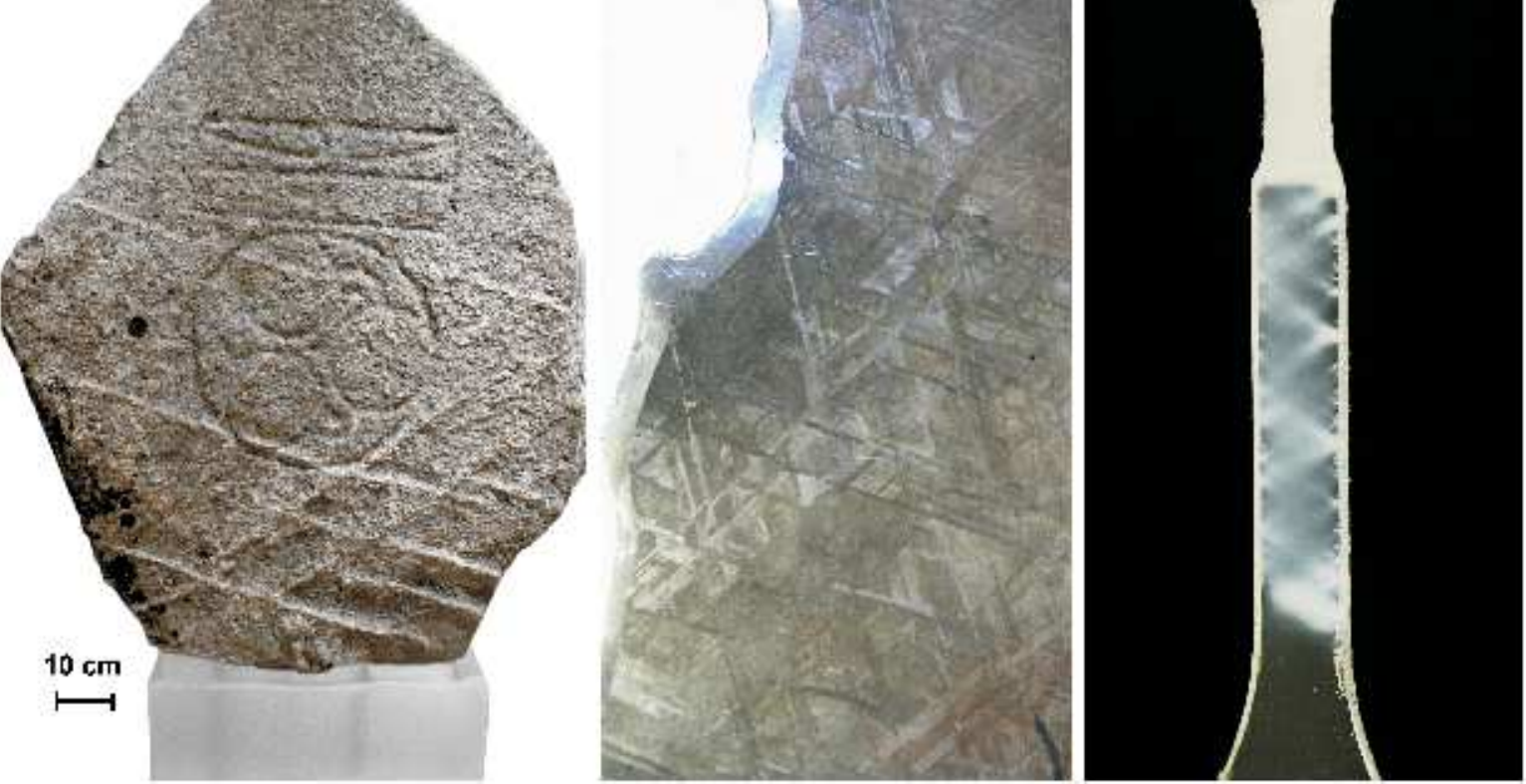} \\
\includegraphics[width=80mm]{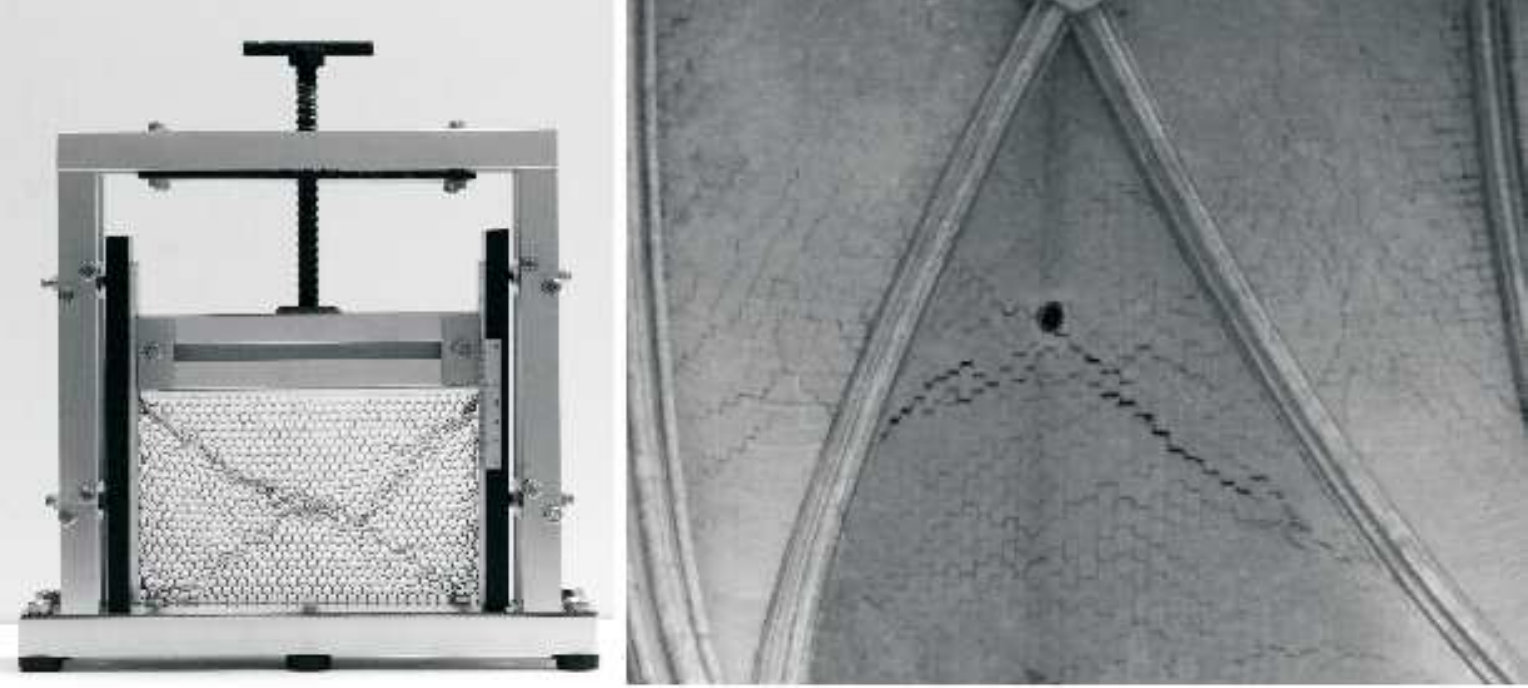} \\
\includegraphics[width=80mm]{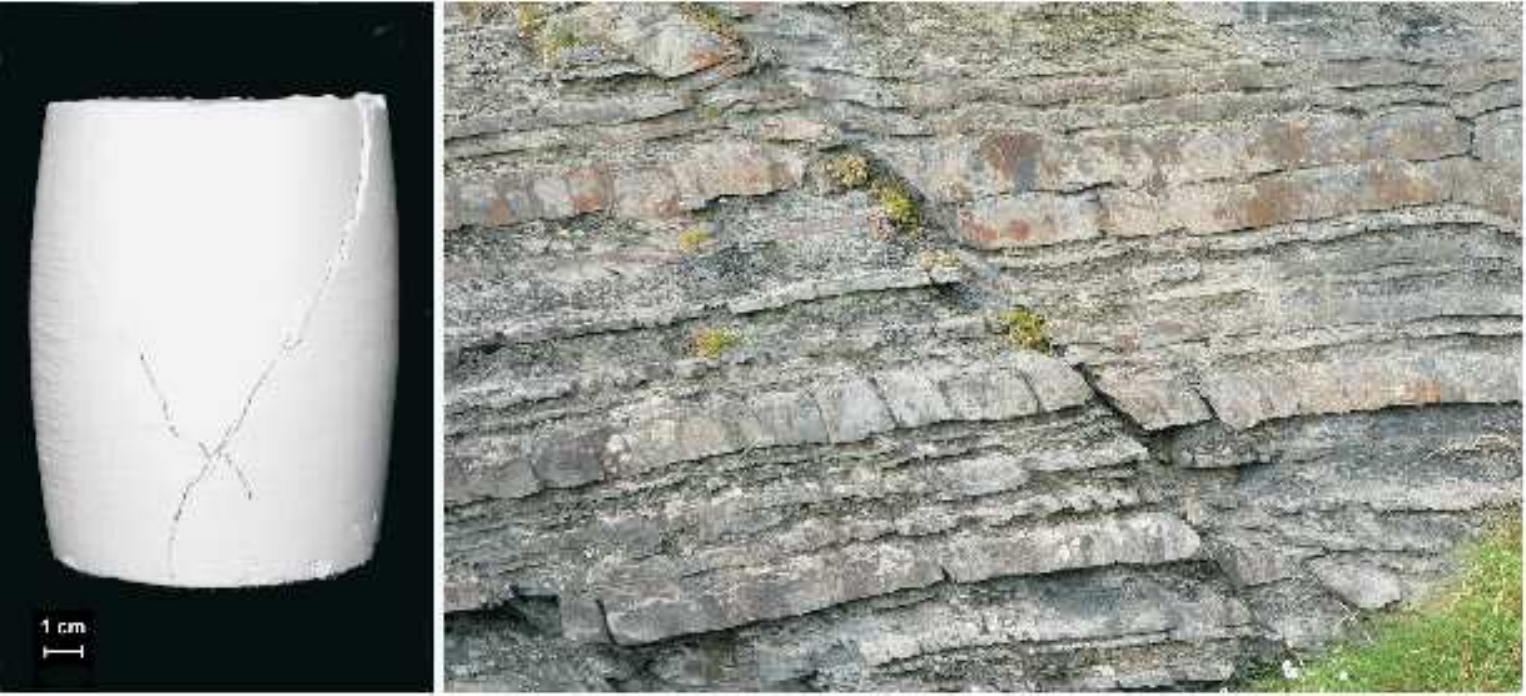} \\
\caption{
\footnotesize
Examples of strain localization. From left to right, starting from the upper part:
A merlon in the Finale Emilia castle failed (during the Emilia earthquake on May 20, 2012) in compression with a typical \lq X-shaped' deformation band pattern (bricks are to be understood here as the  microstructure of a composite material).
A sedimentary rock with the signature of an \lq X-shaped' localization band (infiltrated with a different mineral after formation).
A stone axe from a British Island (Museum of Edinburgh) evidencing two parallel localization bands and another at a different orientation.
A runestone (Museum of Edinburgh) with several localized deformation bands, forming angles of approximatively 45$^\circ$ between each other.
A polished and etched section of an iron meteorite showing several alternate bands of kamacite and taenite.
Deformation bands in a strip of unplasticized poly(vinyl chloride) (uPVC) pulled in tension and eventually evolving into a necking.
An initially regular hexagonal disposition of drinking straws subject to uniform uniaxial strain has evolved into an \lq X-shaped' localization pattern.
A fracture prevails on a regularly distributed network of cracks in a vault of the Amiens dome.
\lq X-shaped' localization bands in a kaolin sample subject to vertical compression and lateral confining pressure.
A thin, isolated localization band in a sedimentary layered rock (Silurian formation near Aberystwyth).
}
\label{fig_a}
\end{figure}

When a ductile material is brought to an extreme strain state through a uniform loading process,
the deformation may start to localize into thin and planar bands, often arranged in regular lattice patterns. This phenomenon is quite common and occurs in many materials over a broad range of scales: from the kilometric scale in the earth crust (Kirby, 1985), down to the nanoscale in metallic glass (Yang, 2005), see the examples reported in Fig.~\ref{fig_a}.

\begin{figure}[!htcb]
\centering
\includegraphics[width=140mm]{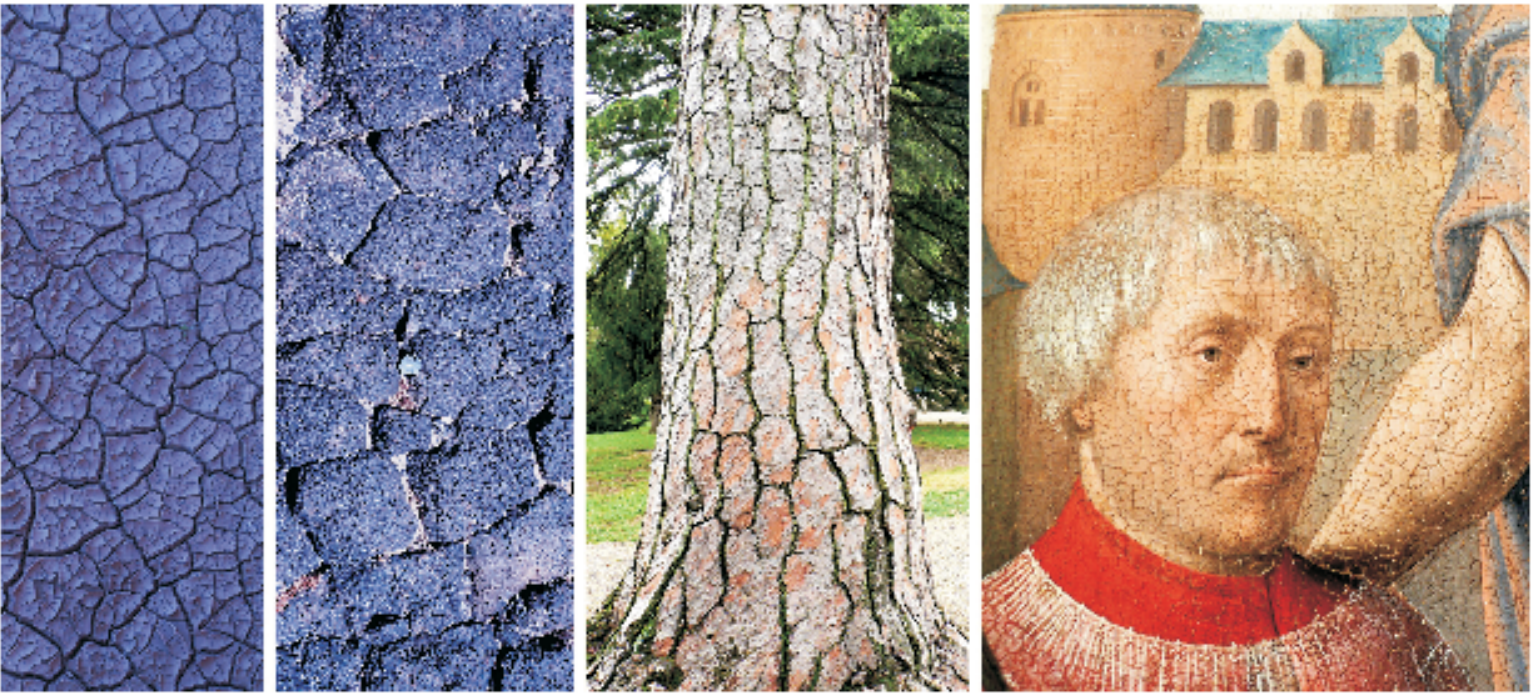}
\caption{\footnotesize
Regular patterns of localized cracks as the signature of strain localization lattices. From left to right:
Dried mud;
Lava cracked during solidification (near Amboy crater);
Bark of a maritime pine (Pinus pinaster);
Cracks in a detail of a painting by J. Provost (\lq Saint Jean-Baptiste', Valenciennes, Mus\'{e}e des Beaux Arts).
}
\label{fig_b}
\end{figure}

After localization, unloading typically\footnote{
For granular materials, there are cases in which unloading occurs inside the shear band, as shown by Gajo et al. (2004).
}
occurs in the material outside the bands, while strain quickly evolves inside, possibly leading to final fracture (as in the examples
shown in Fig.~\ref{fig_b}, where the crack lattice is the signature of the initial shear band network) or to a progressive accumulation of deformation bands
(as for instance in the case of the drinking straws, or of the iron meteorite, or of the uPVC sample shown in Fig.~\ref{fig_a}, or in the well-known case of granular materials, where fracture is usually absent and localization bands are made up of material at a different relative density, Gajo et al. 2004).
%
%

It follows from the above discussion that as strain localization represents a prelude to failure of ductile materials, its mechanical understanding paves the way to the innovative use of materials in extreme mechanical conditions. Although shear bands have been the subject of an intense research effort over the last twenty years (see the review given by Bigoni, 2012), many fundamental questions still remain open: i.) Why are shear bands a preferred mode of failure for ductile materials?
ii.) Why do shear bands propagate rectilinearly under mode II, while cracks do not?
iii.) how does a shear band interact with a crack or with a rigid inclusion? iv.) Does a stress concentration exist at a
shear band tip? v.) How does a shear band behave under dynamic conditions?

The only systematic\footnote{
Special problems of shear band propagation in geological materials have been addressed by
Puzrin and Germanovich (2005) and Rice (1973).
}
attempt to solve these problems seems to have been a series of works by Bigoni and co-workers, based on the perturbative approach
to shear bands (Argani et al. 2014; 2013; Bigoni and Capuani, 2002; 2005; Piccolroaz et al. 2006).
In fact problems (i.), (ii.), and (iv.) were addressed in (Bigoni and Dal Corso, 2008 and Dal Corso and Bigoni, 2010),
problem (iii.) in (Dal Corso et al. 2008; Bigoni et al. 2008; Dal Corso and Bigoni, 2009), and (v.) in (Bigoni and Capuani, 2005).

The purpose of the present article is to present a model of a shear band as a zero-thickness interface and to rigorously motivate this as the asymptotic behaviour of a thin layer of material, which is extremely compliant in shear (Section \ref{sec02}).
Once the shear band model has been developed, it is used (in Section \ref{sec03}) to (i) demonstrate that a shear band grows rectilinearly under mode II remote loading in a material deformed near to failure and (ii.) estimate the stress concentration at the shear band tip.
In particular, a pre-existing shear band is considered to lie in a matrix as a thin zone of material with properties identical to the matrix, but lower yield stress. This is an imperfection, which remains
neutral until the yield is reached in the shear band. 
The present model is based on an imperfection approach and shares similarities to that pursued by Abeyaratne and Triantafyllidis (1981) and Hutchinson and Tvergaard (1981), so that it is essentially
different from a perturbative approach, in which the perturbation is imposed at a certain stage of a uniform deformation process. To highlight the differences and the analogies between the two approaches, the incremental strain field induced by the emergence of a shear band of finite length (modelled as a sliding surface) is determined for a J$_2$-deformation theory material and compared with finite element simulations
in which the shear band is modelled as a zero-thickness layer of compliant material.


\section{Asymptotic model for a thin layer of highly compliant material embedded in a solid}
\label{sec02}

\begin{figure}[!htcb]
\centering
\includegraphics[width=140mm]{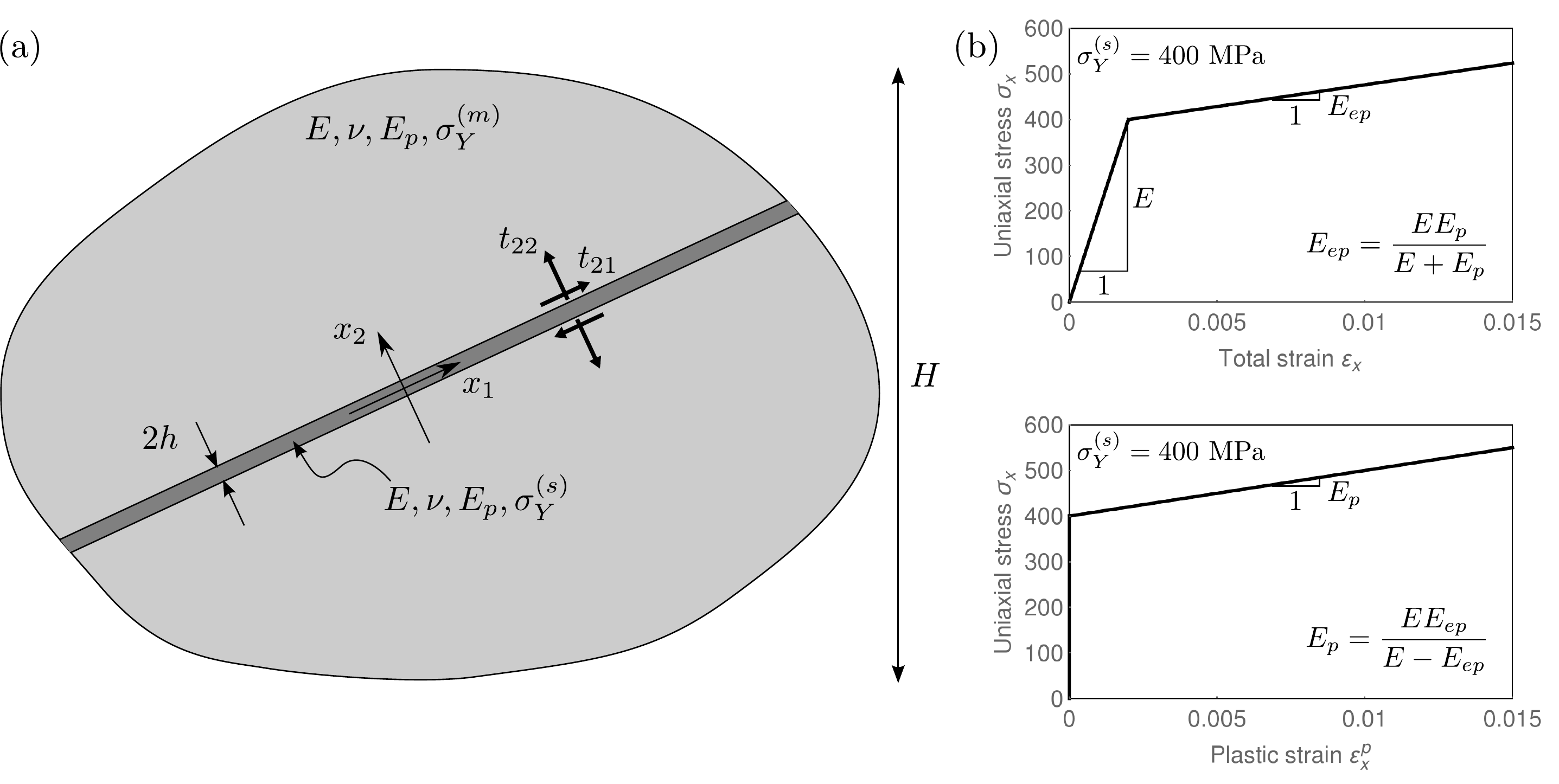}
\caption{(a) A shear band inside a ductile material modeled as a thin layer of highly compliant material ($E_{ep}/E \ll 1$)
embedded in a material block characterized by a dimension $H$, such that $h/H\ll1$;
both materials obey the same von Mises plasticity model represented by the uniaxial stress behaviour reported in (b),
but having a different yield stress (lower inside than outside the shear band).}
\label{fig01}
\end{figure}

A shear band, inside a  solid block of dimension $H$,
is modeled as a thin layer of material (of semi-thickness $h$, with $h/H\ll1$) yielding at a uniaxial stress $\sigma_Y^{(s)}$, which is lower than that of the
surrounding matrix $\sigma_Y^{(m)}$, Fig.~\ref{fig01}.
Except for the yield stress, the material inside and outside the layer is described by the same elastoplastic model,
namely, a von Mises plasticity with associated flow rule and linear hardening,
defined through the elastic constants, denoted by the Young modulus $E$ and Poisson's ratio $\nu$, and the plastic modulus $E_p$, see Fig.~\ref{fig01}b.
%
%

At the initial yielding, the material inside the layer (characterized by a low hardening modulus $E_{ep} = E E_{p}/(E + E_{p})$)
is much more compliant than the material outside (characterized by an elastic isotropic response $E$).

For $h/H \ll 1$, the transmission conditions across the layer imply the continuity of the tractions, $\bt = [t_{21},t_{22}]^T$, which can be expressed in the asymptotic form 
\begin{equation}
\label{cont}
\jump{t_{21}} = O(h), \quad \jump{t_{22}}= O(h),
\end{equation}
where $\jump{\cdot}$ denotes the jump operator.
The jump in displacements, $\jump{\bu} = [\jump{u_1},\jump{u_2}]^T$, across the layer is related to the tractions at its
boundaries through the asymptotic relations (Mishuris et al., 2013; Sonato et al., 2015)
\begin{equation}
\label{trac1}
t_{21}(\jump{u_1},\jump{u_2}) =
\frac{E_p\sqrt{3\jump{u_1}^2 + 4\jump{u_2}^2} + 6h\sigma_Y^{(s)}}{(3E + 2(1 + \nu)E_p)\sqrt{3\jump{u_1}^2 + 4\jump{u_2}^2}} \frac{E \jump{u_1}}{2h} + O(h),
\end{equation}
\begin{equation}
\label{trac2}
t_{22}(\jump{u_1},\jump{u_2}) =
\frac{(E + 2(1 - \nu)E_p)\sqrt{3\jump{u_1}^2 + 4\jump{u_2}^2} + 8h(1 - 2\nu)\sigma_Y^{(s)}}{(1 - 2\nu)(3E + 2(1 + \nu)E_p)\sqrt{3\jump{u_1}^2 + 4\jump{u_2}^2}} \frac{E \jump{u_2}}{2h} + O(h),
\end{equation}
involving the semi-thickness $h$ of the shear band, which enters the formulation as
a {\it constitutive parameter for the zero-thickness interface model} and introduces a {\it length scale}.
Note that, by neglecting the remainder $O(h)$, Eqs.~(\ref{trac1}) and (\ref{trac2}) define nonlinear relationships between tractions and jump in displacements.

The time derivative of Eqs.~(\ref{trac1}) and (\ref{trac2}) yields the following asymptotic relation between incremental quantities 
\begin{equation}
\label{eq:incr}
\dot\bt \sim \left[\frac{1}{h}\bK_{-1} + \bK_0(\jump{u_1},\jump{u_2}) \right]\jump{\dot\bu},
\end{equation}
where the two stiffness matrices $\bK_{-1}$ and $\bK_{0}$ are given by
\begin{align}
\bK_{-1} &= \ds
\frac{E}{2 (3E + 2(1 + \nu)E_p)}
\begin{bmatrix}
\ds E_p&
\ds 0 \\
\ds 0 &
\ds \frac{E + 2(1 - \nu)E_p}{1 - 2\nu}
\end{bmatrix},
\\[3mm]
\bK_{0} &= \ds
\frac{12  E \sigma_Y^{(s)} }{(3E + 2(1 + \nu)E_p) (3\jump{u_1}^2 + 4\jump{u_2}^2)^{3/2}}
\begin{bmatrix}
\ds  \jump{u_2}^2 &
\ds - \jump{u_1}\jump{u_2} \\
\ds -\jump{u_1}\jump{u_2} &
\ds \jump{u_1}^2
\end{bmatrix},
\end{align}
Assuming now a perfectly plastic behaviour, $E_p=0$, in the limit $h/H\rightarrow 0$ the condition
\beq
\jump{ u_2} = 0
\eeq
is obtained, so that the incremental transmission conditions (\ref{eq:incr}) can be approximated to the leading order as
\begin{equation}
\label{eq:incr0}
\dot\bt \sim \frac{1}{h}\bK_{-1} \jump{\dot\bu}.
\end{equation}
Therefore, when the material inside the layer is close to the perfect plasticity condition, the  incremental conditions assume the limit value
\begin{equation}
\label{shearb}
\dot t_{21} = 0, \quad \jump{\dot u_2} = 0,
\end{equation}
which, together with the incremental version of eq. (\ref{cont})$_2$, namely,
\begin{equation}
\label{shearb2}
\quad \jump{\dot{t}_{22}} = 0,
\end{equation}
correspond to the incremental boundary conditions proposed in Bigoni and Dal Corso (2008) to define a pre-existing shear band of null thickness.

The limit relations (\ref{shearb}) and (\ref{shearb2}) motivate the use of the imperfect interface approach (Mishuris, 2004; Mishuris and Ochsner, 2005; 2007; Antipov et al. 2001;  Bigoni et al., 1998) for the modelling of shear band growth in a ductile material.
A computational model, in which the shear bands are modelled as interfaces, is presented in the next section.

\section{Numerical simulations}
\label{sec03}

Two-dimensional plane-strain finite element simulations are presented to show the effectiveness of the above-described asymptotic model for a
thin and highly compliant layer in modelling a shear band embedded in a ductile material. Specifically,
we will show that the model predicts rectilinear propagation of a shear band under simple shear boundary conditions and
it allows the investigation of the stress concentration at the shear band tip.

\begin{figure}[!htcb]
\centering
\includegraphics[width=90mm]{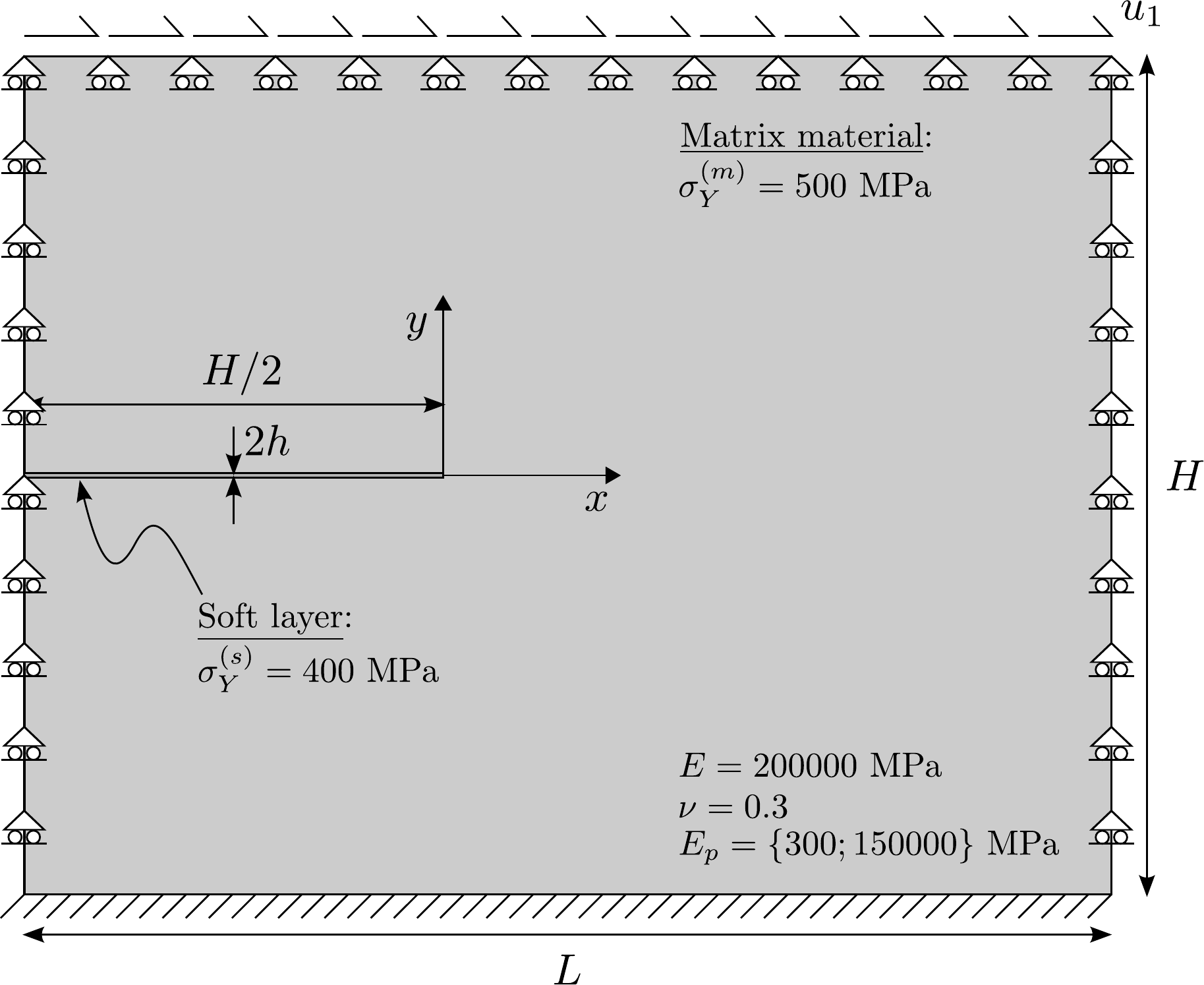}
\caption{Geometry of the model, material properties and boundary conditions (that would correspond to a simple shear deformation in the absence of the shear band).
The horizontal displacement $u_1$ is prescribed at the upper edge of the domain.}
\label{fig02}
\end{figure}

The geometry and material properties of the model are shown in Fig.~\ref{fig02}, where a rectangular block of edges $H$ and $L\geq H$
is subject to boundary conditions consistent with a simple shear deformation, so that the lower edge of the square domain is clamped,
the vertical displacements are constrained along the vertical edges and along the upper edge, where a constant horizontal displacement $u_1$ is prescribed.
The domain is made of a ductile material and contains a thin ($h/H \ll 1$) and highly compliant ($E_{ep}/E\ll1$) layer of length $H/2$ and thickness $2h = 0.01$ mm,
 which models a shear band.
The material constitutive behaviour is described by an elastoplastic model based on linear isotropic elasticity ($E=200000$ MPa, $\nu=0.3$)
and von Mises plasticity with linear hardening (the plastic modulus is denoted by $E_p$).
The uniaxial yield stress $\sigma_Y^{(m)}$ for the matrix material is equal to 500 MPa, whereas the layer is characterized by a lower yield stress, namely,
$\sigma_Y^{(s)} = 400$ MPa.
%

The layer remains neutral until yielding, but, starting from that stress level, it becomes a material inhomogeneity, being more compliant (because its response is characterized by $E_{ep}$) than the matrix (still in the elastic regime and thus characterized by $E$). The layer can be representative of a pre-existing shear band and can be treated with the zero-thickness interface model, Eqs.~(\ref{trac1}) and (\ref{trac2}).
This zero-thickness interface was implemented in the ABAQUS finite element software\footnote{ABAQUS Standard Ver. 6.13 has been used, available on the AMD Opteron cluster
Stimulus at UniTN.} through cohesive elements, equipped with the traction-separation laws, Eqs.~(\ref{trac1}) and (\ref{trac2}), by means of the user subroutine UMAT.
An interface, embedded into the cohesive elements, is characterized by two dimensions: a geometrical and a constitutive thickness. The latter, $2h$, exactly corresponds to the constitutive thickness involved in the model for the interface (\ref{trac1}) and (\ref{trac2}), while the former, denoted by $2h_g$, is related to the mesh dimension in a way that the results become independent of this parameter, in the sense that a mesh refinement yields results converging to a well-defined solution.

We consider two situations. In the first, we assume that the plastic modulus is $E_p = 150000$ MPa (both inside and outside the shear band), so that the material is in a state far from a shear band instability (represented by loss of ellipticity of the tangent constitutive operator, occurring at $E_p = 0$) when at yield.
In the second, we assume that the material is prone to a shear band instability, though still in the elliptic regime, so that $E_p$ (both inside and outside the shear band) is selected to be \lq sufficiently small', namely, $E_p = 300$ MPa. 
The pre-existing shear band is therefore employed as an imperfection triggering shear strain localization when the material is still inside the region, but close to the boundary, of ellipticity.


\subsection{Description of the numerical model}

\begin{figure}[!htcb]
\centering
\includegraphics[width=50mm]{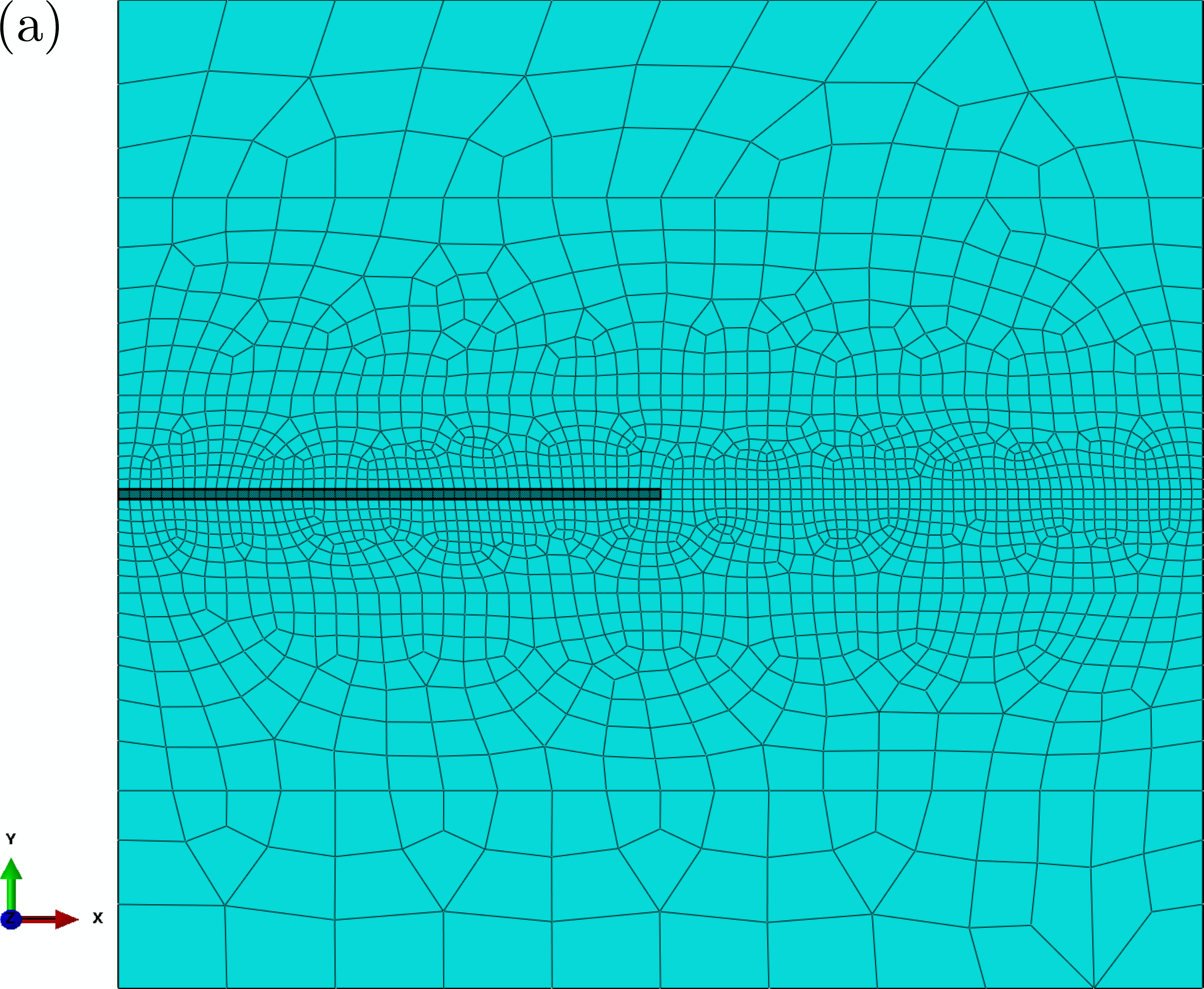}
\includegraphics[width=50mm]{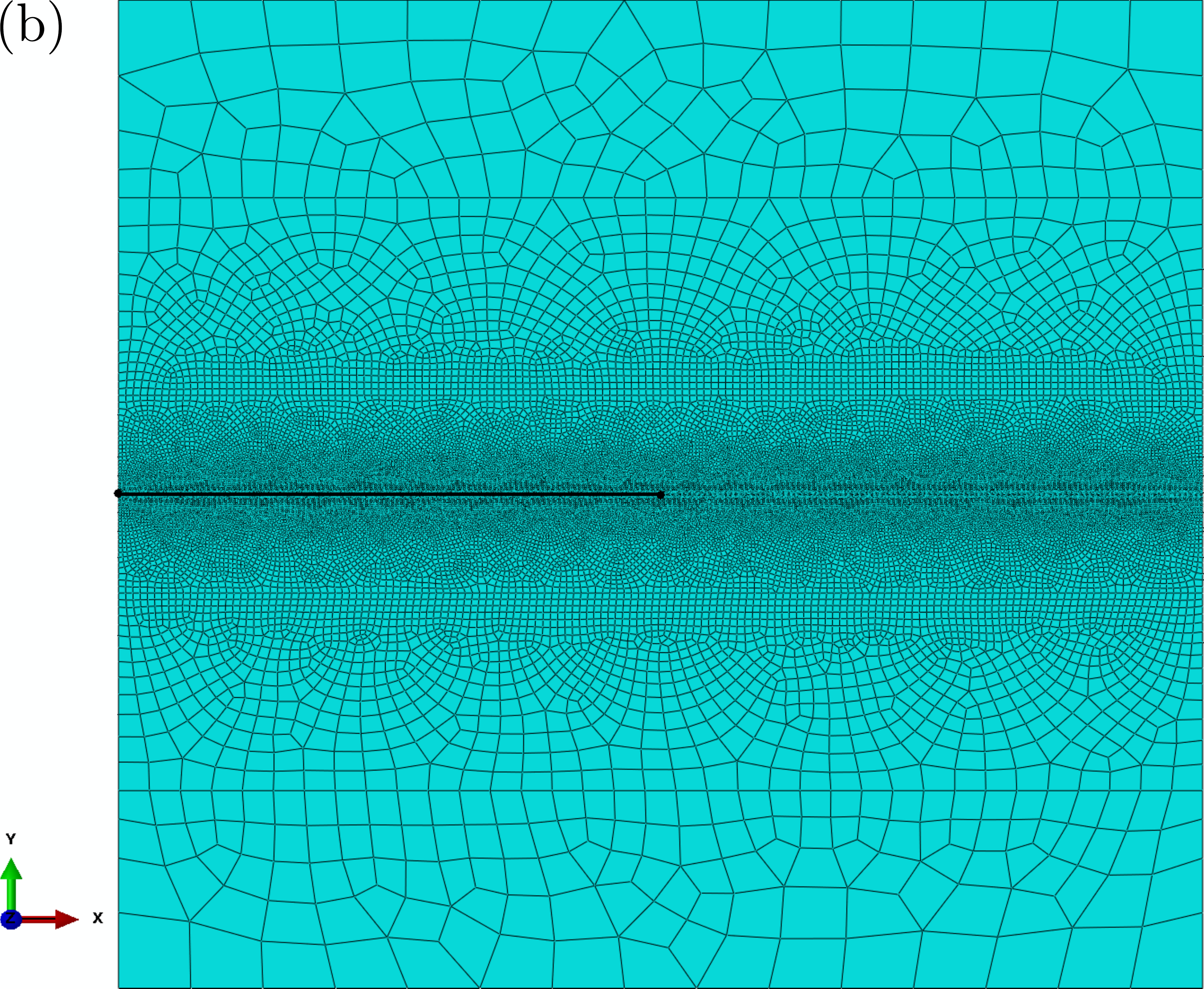}
\includegraphics[width=50mm]{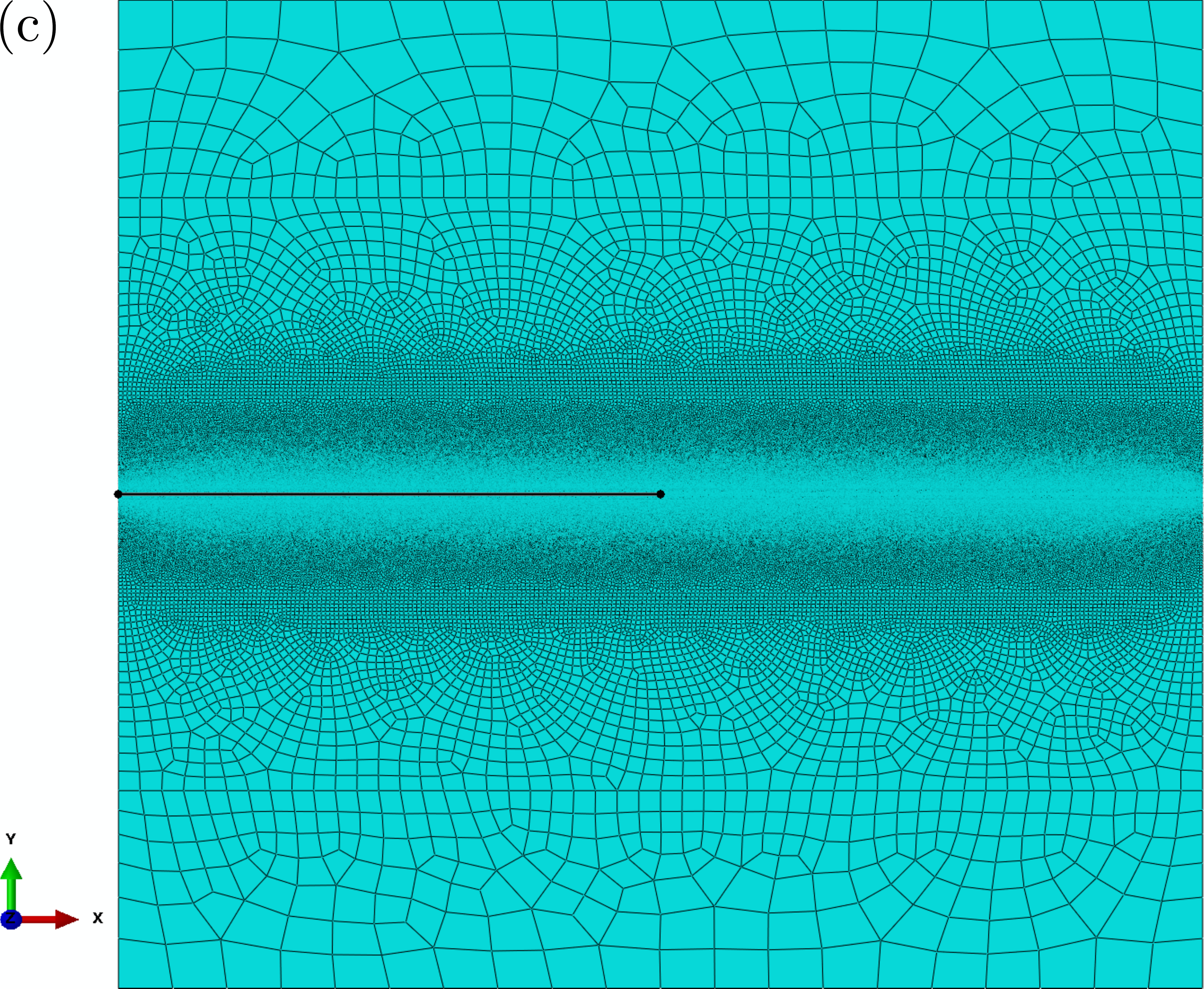}
\caption{The three meshes used in the analysis to simulate a shear band (highlighted in black) in a square solid block ($L=H=$ 10 mm).
The shear band is represented in the three cases as an interface with the same constitutive thickness $h=0.005$ mm,
but with decreasing geometric thickness $h_g$; (a) coarse mesh (1918 nodes, 1874 elements, $h_g$ = 0.05 mm);
(b) fine mesh (32079 nodes, 31973 elements, $h_g$ = 0.005 mm);
(c) ultra-fine mesh (1488156 nodes, 1487866 elements, $h_g$ = 0.0005 mm)}
\label{mesh}
\end{figure}

With reference to a square block ($L=H=$ 10 mm) containing a pre-existing shear band with constitutive thickness $h=$ 0.005 mm, three different meshes were used, differing in the geometrical
thickness of the interface representing the pre-existing shear band (see Fig.~\ref{mesh} where the shear band is highlighted with a black line),
namely, $h_g = \{0.05; 0.005; 0.0005\}$ mm corresponding to coarse, fine, and ultra-fine meshes.

The three meshes were generated automatically using the mesh generator available in ABAQUS. In order to have increasing mesh refinement from the exterior (upper and lower parts) to the interior (central part) of the domain, where the shear band is located, and to ensure the appropriate element shape and size according to the geometrical thickness $2h_g$, the domain was partitioned into rectangular subdomains with increasing mesh seeding from the exterior to the interior. Afterwards, the meshes were generated by employing a free meshing technique with quadrilateral elements and the advancing front algorithm.

The interface that models the shear band is discretized using 4-node two-dimensional cohesive elements (COH2D4), while the matrix material is modelled using 4-node bilinear, reduced integration with hourglass control (CPE4R).

It is important to note that the constitutive thickness used for traction-separation response is always equal to the actual size of the shear band $h = 0.005$ mm, whereas the geometric thickness $h_g$, defining the height of the cohesive elements, is different for the three different meshes. 
Consequently, all the three meshes used in the simulations correspond to the same problem in terms of both material properties and geometrical dimensions (although the geometric size of the interface is different),
so that the results have to be, and indeed will be shown to be, mesh independent.

\subsection{Numerical results}

\begin{figure}[!htcb]
\centering
\includegraphics[width=140mm]{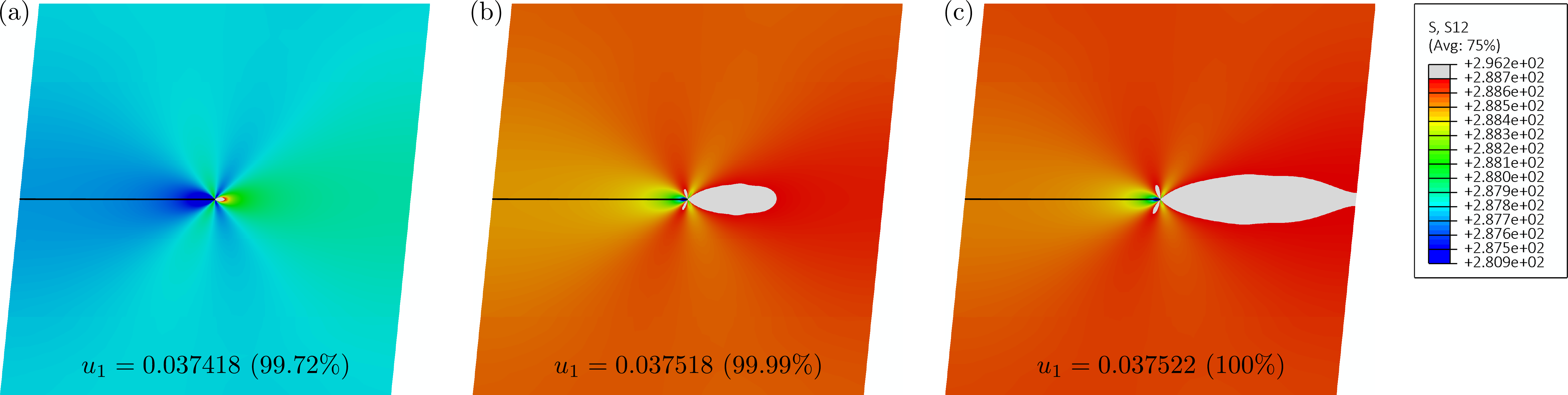}
\caption{Contour plots of the shear stress $\sigma_{12}$ for the case of material far from shear band instability ($E_p = 150000$ MPa). The grey region corresponds to the material at yielding $\sigma_{12} \geq 500/\sqrt{3} \backsimeq 288.68$ MPa. Three different stages of deformation are shown, corresponding to a prescribed displacement at the upper edge of the square domain $u_1 = 0.037418$ mm (a), $u_1 = 0.037518$ mm (b), $u_1 = 0.037522$ mm (c). The displacements in the figures are amplified by a deformation scale factor of 25 and the percentages refer to the final displacement.}
\label{contour_s12_case3}
\end{figure}

\begin{figure}[!htcb]
\centering
\includegraphics[width=140mm]{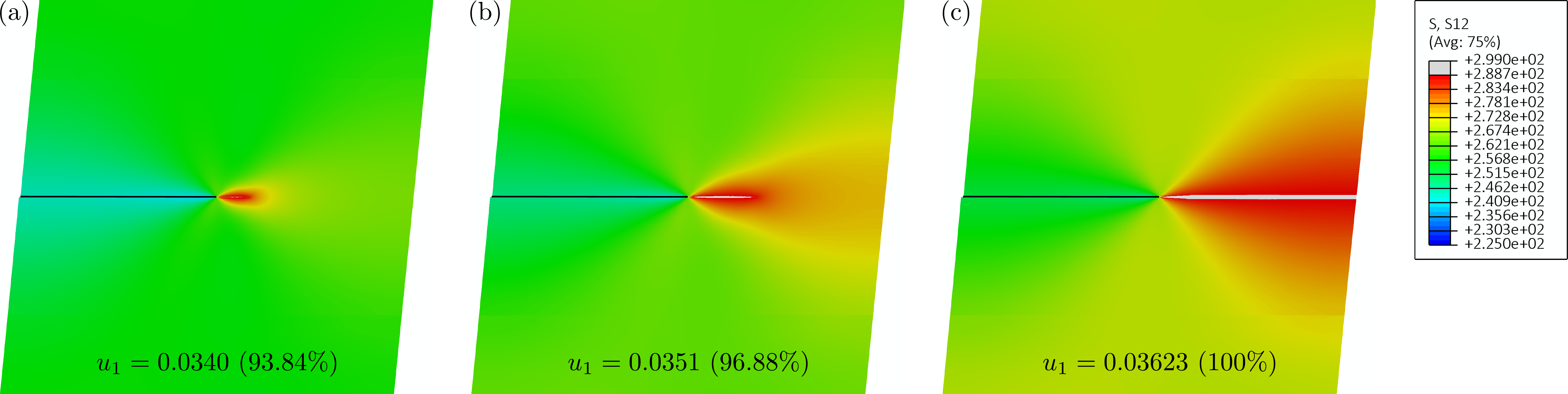}
\caption{Contour plots of the shear stress $\sigma_{12}$ for the case of material close to shear band instability ($E_p = 300$ MPa). The grey region corresponds to the material at yielding $\sigma_{12} \geq 500/\sqrt{3}$. Three different stages of deformation are shown, corresponding to a prescribed displacement at the upper edge of the square domain $u_1 = 0.0340$ mm (a), $u_1 = 0.0351$ mm (b), $u_1 = 0.03623$ mm (c). The displacements in the figures are amplified by a deformation scale factor of 27.}
\label{contour_s12_case1}
\end{figure}

Results (obtained using the fine mesh, Fig.~\ref{mesh}b) in terms of the shear stress component $\sigma_{12}$ at different stages of a deformation process for the boundary value problem sketched in Fig.~\ref{fig02} are reported in Figs.~\ref{contour_s12_case3} and \ref{contour_s12_case1}. 

In particular, Fig.~\ref{contour_s12_case3} refers to a matrix with high plastic modulus, $E_p = 150000$ MPa, so that the material is 
far from the shear band formation threshold. 
The upper limit of the contour levels was set to the value $\sigma_{12} = 500/\sqrt{3} \backsimeq 288.68$ MPa, corresponding to the yielding stress of the matrix material. As a result, the grey zone in the figure represents the material at yielding, whereas the material outside the grey zone is still in the elastic regime.
Three stages of deformation are shown, corresponding to: the initial yielding  of the matrix material (left), the yielding zone occupying approximately one half of the 
space between the shear band tip and the right edge of the domain
(centre), and the yielding completely linking the tip of the shear band to the boundary (right).
Note that the shear band, playing the role of a material imperfection, produces a stress concentration at its tip. However, the region of high stress level rapidly grows and diffuses in the whole domain. At the final stage, shown in Fig.~\ref{contour_s12_case3}c, almost all the matrix material is close to yielding.

Fig.~\ref{contour_s12_case1} refers to a matrix with low plastic modulus, $E_p = 300$ MPa, so that the 
material is close (but still in the elliptic regime) to the shear band formation threshold ($E_p = 0$). 
Three stages of deformation are shown, from the condition of initial yielding of the matrix material near the shear band tip  (left), to an intermediate condition (centre), and finally to the complete yielding of a narrow zone connecting the shear band tip to the right boundary (right).
In this case, where the material is prone to shear band localization, the zone of high stress level departs from the shear band tip and propagates towards the right. This propagation occurs in a highly concentrated narrow layer, rectilinear, and parallel to the pre-existing shear band. At the final stage of deformation, shown in Fig.~\ref{contour_s12_case1}c, the layer of localized shear has reached the boundary of the block.

\begin{figure}[!htcb]
\centering
\includegraphics[width=140mm]{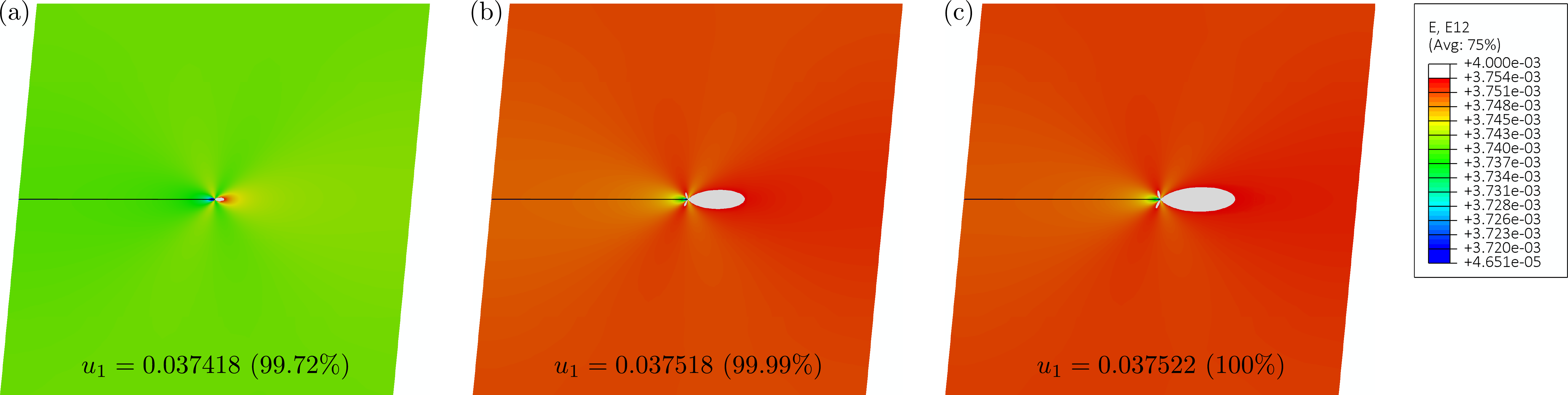}
\caption{Contour plots of the shear deformation $\gamma_{12}$ for the case of material far from shear band instability ($E_p = 150000$ MPa). Three different stages of deformation are shown, corresponding to a prescribed displacement at the upper edge of the square domain $u_1 = 0.037418$ mm (a), $u_1 = 0.037518$ mm (b), $u_1 = 0.037522$ mm (c). The displacements in the figures are amplified by a deformation scale factor of 25.}
\label{contour_e12_case3}
\end{figure}

\begin{figure}[!htcb]
\centering
\includegraphics[width=140mm]{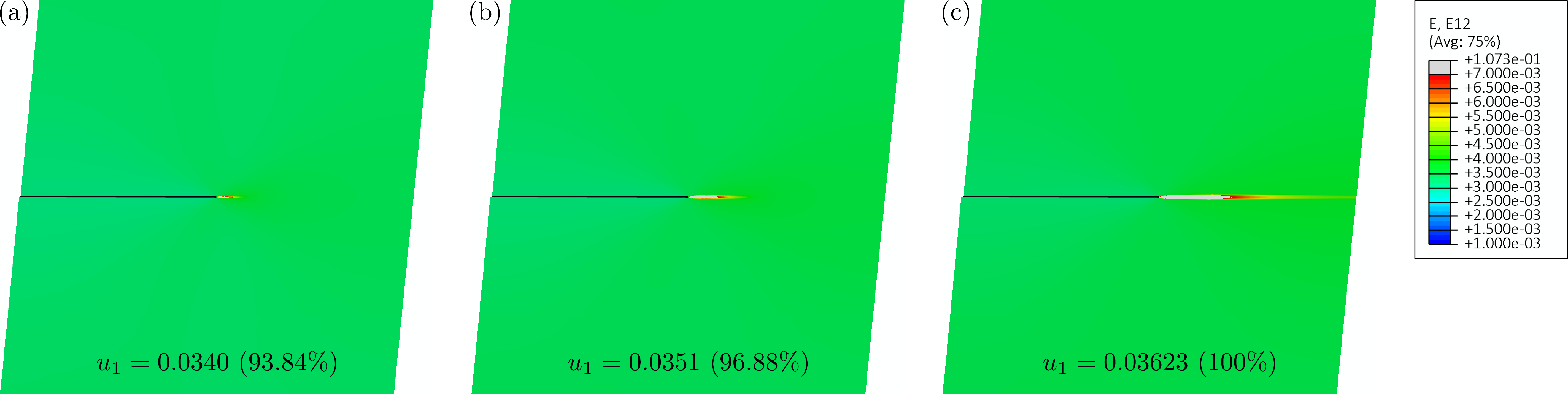}
\caption{Contour plots of the shear deformation $\gamma_{12}$ for the case of material close to shear band instability ($E_p = 300$ MPa). Three different stages of deformation are shown, corresponding to a prescribed displacement at the upper edge of the square domain $u_1 = 0.0340$ mm (a), $u_1 = 0.0351$ mm (b), $u_1 = 0.03623$ mm (c). The displacements in the figures are amplified by a deformation scale factor of 27.}
\label{contour_e12_case1}
\end{figure}

Results in terms of the shear strain component $\gamma_{12}$, for both cases of material far from, and close to shear band instability are reported in Figs.~\ref{contour_e12_case3} and \ref{contour_e12_case1}, respectively.
In particular, Fig.~\ref{contour_e12_case3} shows contour plots of the shear deformation $\gamma_{12}$ for the case of a material far from the shear band instability ($E_p = 150000$ MPa) at the same three stages of deformation as those reported in Fig.~\ref{contour_s12_case3}. Although the tip of the shear band acts as a strain raiser, the contour plots show that the level of shear deformation is high and remains diffused in the whole domain.

Fig.~\ref{contour_e12_case1} shows contour plots of the shear deformation $\gamma_{12}$ for the case of a material close to the shear band instability ($E_p = 300$ MPa), at the same three stages of deformation as  those reported in Fig.~\ref{contour_s12_case1}. It is noted that the shear deformation is localized along a rectilinear path ahead of the shear band tip, confirming results that will be reported 
later with the  perturbation approach (Section \ref{sec:pert}).

\begin{figure}[!htcb]
\centering
\includegraphics[width=160mm]{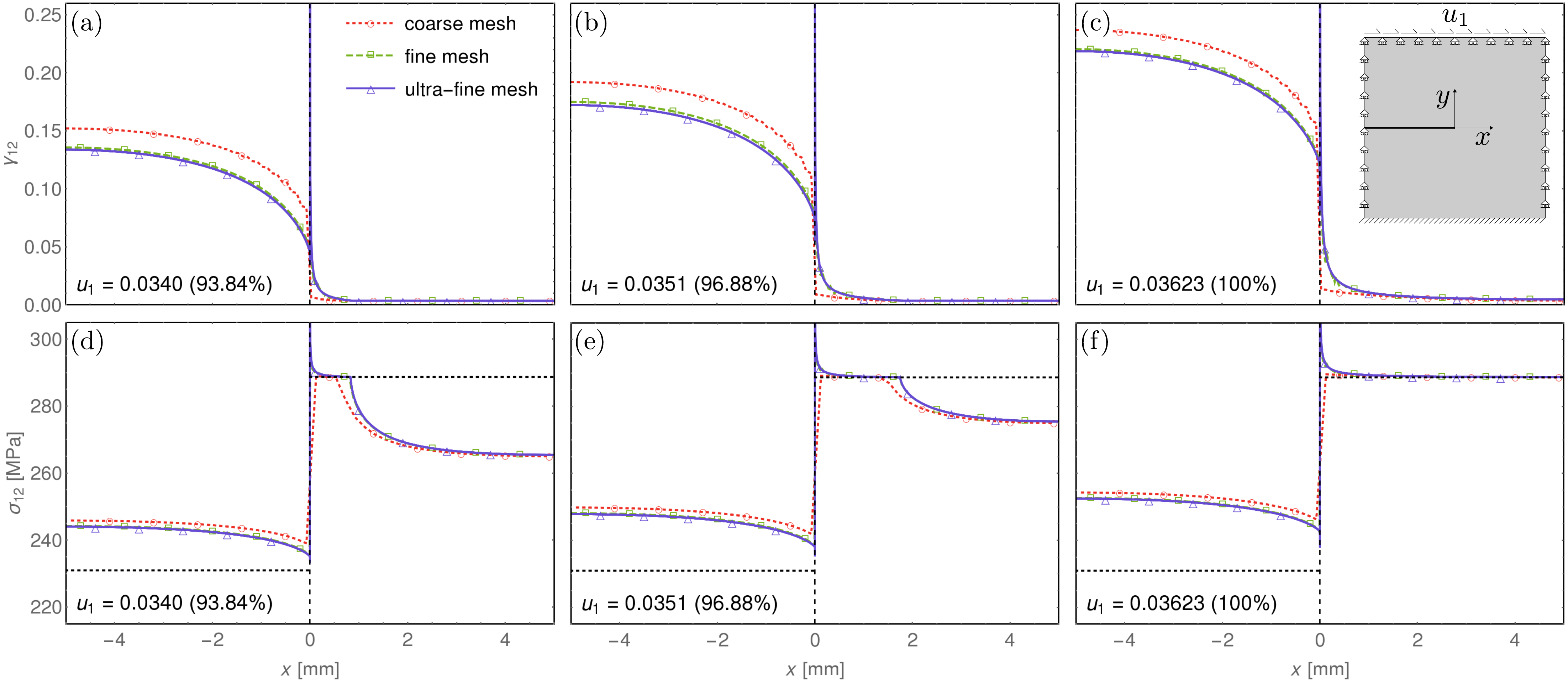}
\caption{Shear deformation $\gamma_{12}$ (upper part) and shear stress $\sigma_{12}$ (lower part) along the $x$-axis containing the pre-existing shear band for the case of a material close to a shear band instability $E_p = 300$ MPa. The black dotted line, in the bottom part of the figure, indicates the yield stress level, lower inside the pre-existing shear band than that in the outer domain. Three different stages of deformation are shown, corresponding to a prescribed displacement at the upper edge of the square domain $u_1 = 0.0340$ mm (left), $u_1 = 0.0351$ mm (center), $u_1 = 0.03623$ mm (right).}
\label{pathcase1}
\end{figure}

The shear deformation $\gamma_{12}$ and the shear stress $\sigma_{12}$ along the $x$-axis containing the pre-existing shear band for the case of a material close to strain localization, $E_p = 300$ MPa, are shown in Fig.~\ref{pathcase1}, upper and lower parts, respectively. Results are reported for the three meshes, coarse, fine and ultra-fine (Fig.~\ref{mesh}) and at the same three stages of deformation as those shown in  Figs.~\ref{contour_s12_case1} and \ref{contour_e12_case1}. The results appear to be mesh independent, meaning that the solution converges as the mesh is more and more refined. 

The deformation process reported in Figs.~\ref{contour_s12_case1}, \ref{contour_e12_case1}, and \ref{pathcase1} can be described as follows. After an initial homogeneous elastic deformation (not shown in the figure), in which the shear band remains neutral (since it shares the same elastic properties with the matrix material), the stress level reaches $\sigma_{12} = 400/\sqrt{3} \backsimeq 230.9$ MPa, corresponding to the yielding of the material inside the shear band. Starting from this point, the pre-existing shear band is activated, which is confirmed by a high shear deformation $\gamma_{12}$ and a stress level above the yield stress inside the layer, $-5 \text{ mm} < x < 0$ (left part of Fig. \ref{pathcase1}). The activated shear band induces a strain localization and a stress concentration at its tip, thus generating a zone of material at yield, which propagates to the right (central part of Fig. \ref{pathcase1}) until collapse (right part of Fig. \ref{pathcase1}).

\begin{figure}[!htcb]
\centering
\includegraphics[width=160mm]{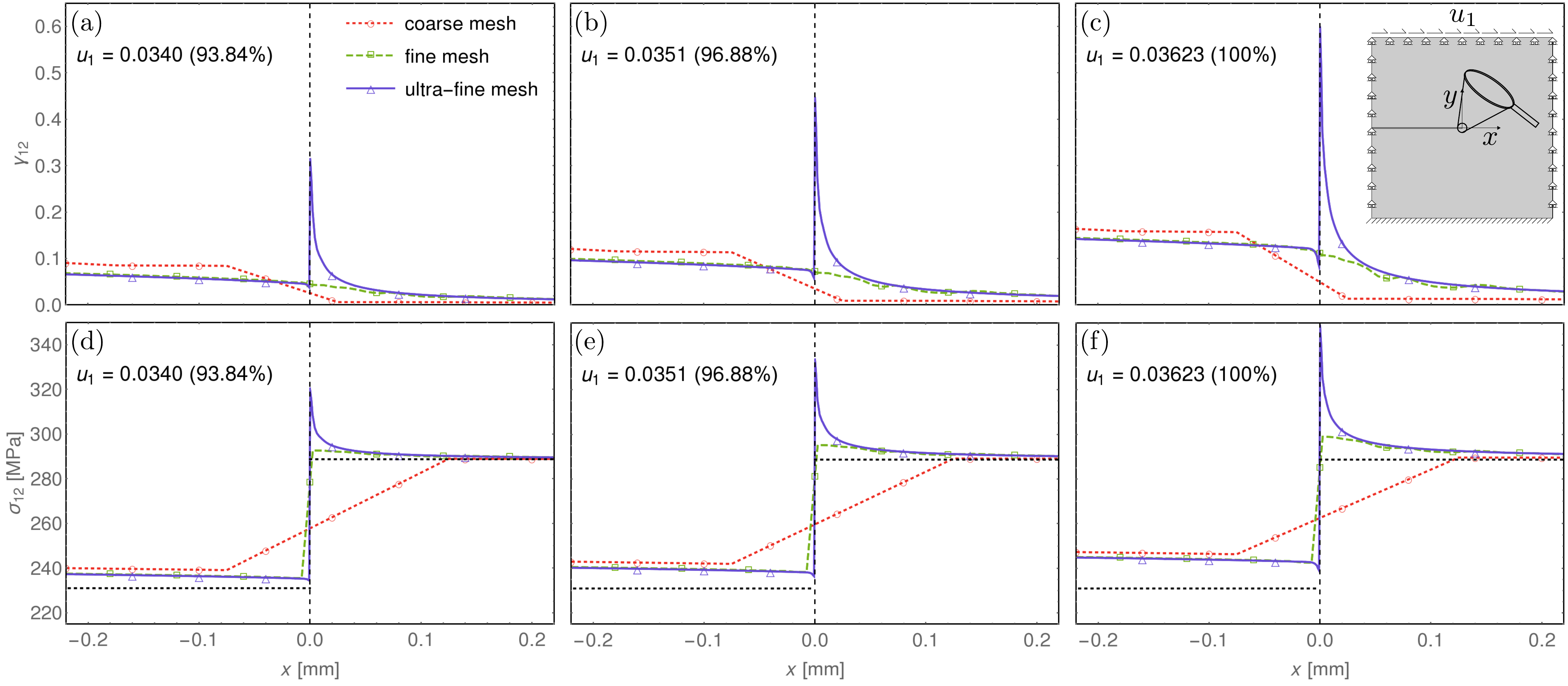}
\caption{Shear and stress concentration at the shear band tip.
Shear deformation $\gamma_{12}$ (upper part) and shear stress $\sigma_{12}$ (lower part) along the $x$-axis containing the pre-existing shear band for the case of a material close to a shear band instability $E_p = 300$ MPa. Three different stages of deformation are shown, corresponding to a prescribed displacement at the upper edge of the square domain $u_1 = 0.0340$ mm (left), $u_1 = 0.0351$ mm (center), $u_1 = 0.03623$ mm (right).}
\label{pathcase1Z}
\end{figure}

In order to appreciate the strain and stress concentration at the shear band tip, a magnification of the results shown in 
Fig. \ref{pathcase1}
in the region $-0.2 \text{ mm} < x < 0.2 \text{ mm}$ is presented in Fig.~\ref{pathcase1Z}.
Due to the strong localization produced by the shear band, only the ultra-fine mesh is able to capture accurately the strain and stress raising (blue solid curve), whereas the coarse and fine meshes smooth over the strain and stress levels (red dotted and green dashed curves, respectively).
The necessity of a ultra-fine mesh to capture details of the stress/strain fields is well-known in computational fracture mechanics, where special elements (quarter-point or extended elements) have been 
introduced to avoid the use of these ultra-fine meshes at corner points.

\begin{figure}[!htcb]
\centering
\includegraphics[width=160mm]{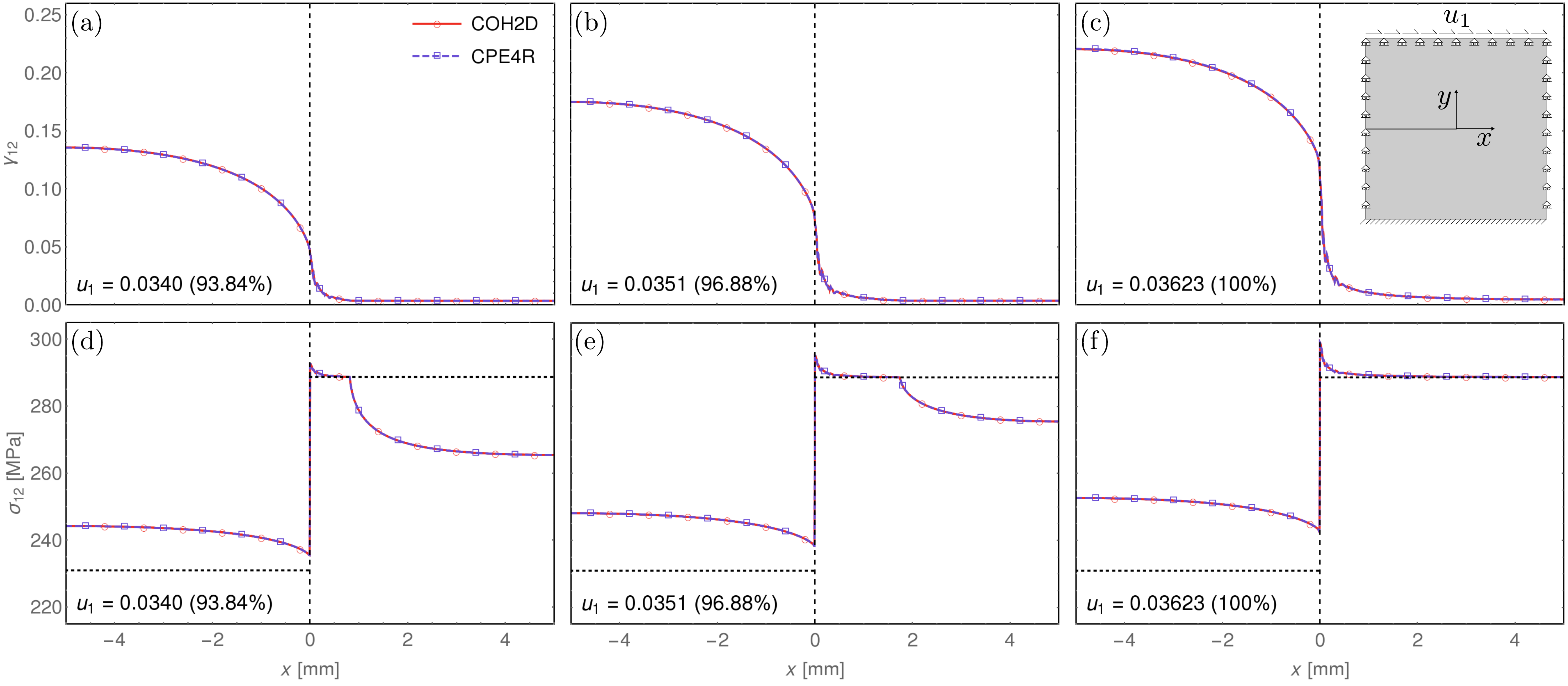}
\caption{Results of simulations performed with different idealizations for the shear band: zero-thickness model (discretized with cohesive elements, COH2D) versus a true layer description (discretized with CPE4R elements).
Shear deformation $\gamma_{12}$ (upper part) and shear stress $\sigma_{12}$ (lower part) along the horizontal line $y = 0$ containing the pre-existing shear band for the case of a material close to a shear band instability $E_p = 300$ MPa. Three different stages of deformation are shown, corresponding to a prescribed displacement at the upper edge of the square domain $u_1 = 0.0340$ mm (left), $u_1 = 0.0351$ mm (center), $u_1 = 0.03623$ mm (right).}
\label{pathcase1b}
\end{figure}

For the purpose of a comparison with an independent and fully numerical representation of the shear band, a finite element simulation was also been performed, using standard continuum elements (CPE4R) instead of cohesive elements (COH2D4) inside the layer. This simulation is important to assess the validity of the asymptotic model of the layer presented in Sec. \ref{sec02}.
In this simulation, reported in Fig.~\ref{pathcase1b}, the layer representing the shear band is a \lq true' layer of a given and finite thickness, thus influencing the results (while these are independent of the geometrical thickness $2h_g$ of the cohesive elements, when the constitutive thickness $2h$ is the same). 
Therefore,  only the fine mesh, shown in Fig.~\ref{mesh}b, was used, as it corresponds to the correct size of the shear band. 
The coarse mesh (Fig.~\ref{mesh}a) and the ultra-fine mesh (Fig.~\ref{mesh}c)  would obviously produce different results, corresponding respectively to a thicker or thinner layer.
Results pertaining to the asymptotic model, implemented into the traction-separation law for the cohesive elements COH2D, are also reported in the figure (red solid curve) and are spot-on with the results obtained with a fully numerical solution employing standard continuum elements CPE4R (blue dashed curve).

\begin{figure}[!htcb]
\centering
\includegraphics[width=160mm]{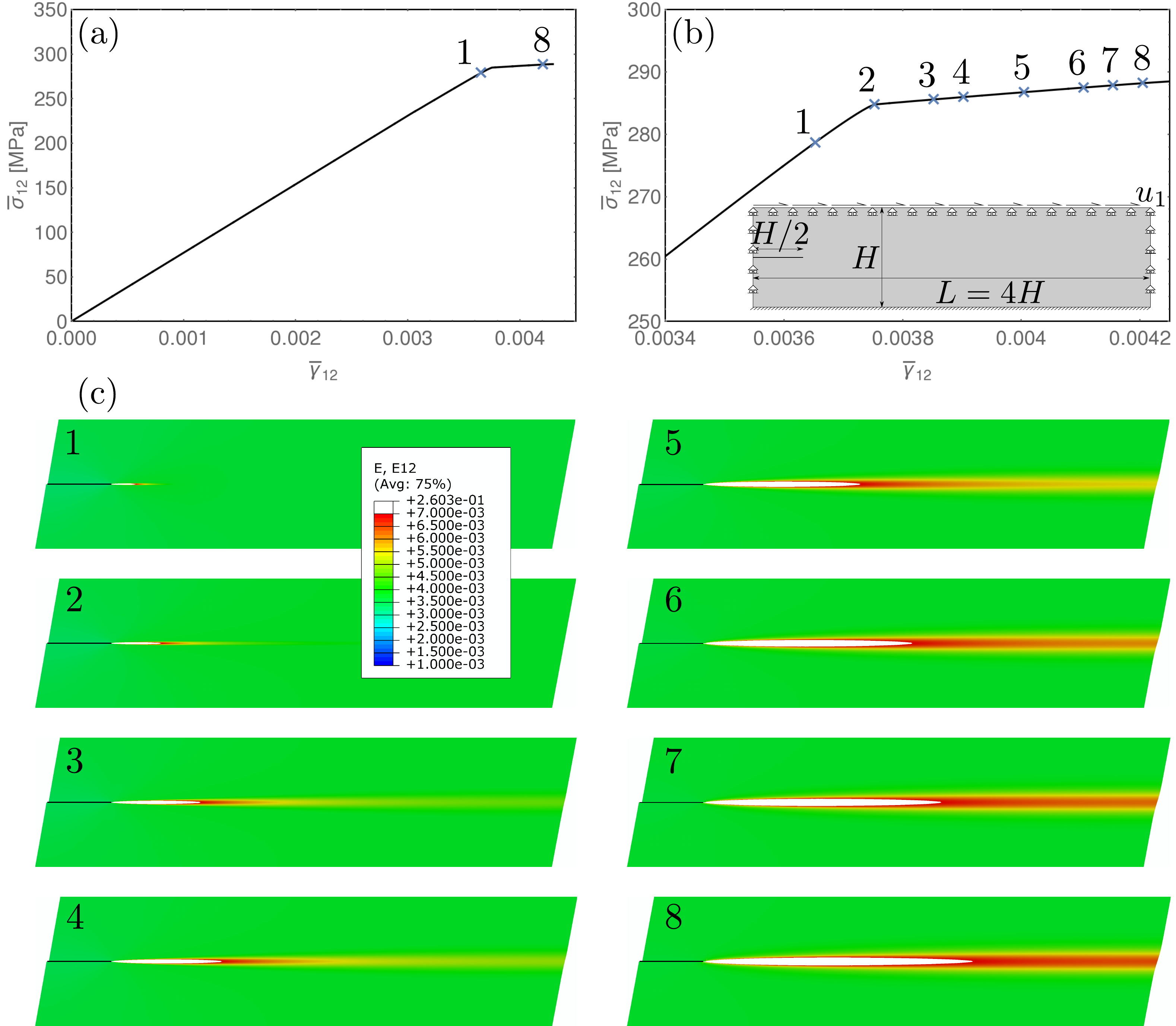}
\caption{Results for a rectangular domain ($L = 40$ mm, $H = 10$ mm) of material close to shear band instability ($E_p = 300$ MPa) and containing a preexisting shear band (of length $H/2 = 5$ mm and constitutive thickness $2h = 0.01$ mm). (a) Overall response curve of the block in terms of average shear stress $\bar{\sigma}_{12} = T/L$, where $T$ is the total shear reaction force at the upper edge of the block, and average shear strain $\bar{\gamma}_{12} = u_1/H$. (b) Magnification of the overall response curve $\bar{\sigma}_{12} - \bar{\gamma}_{12}$ around the stress level corresponding to the yielding of the shear band.
(c) Contour plots of the shear deformation $\gamma_{12}$ at different stages of deformation, corresponding to the points along the overall response curve shown in part (b) of the figure. The deformation is highly focused along a rectilinear path emanating from the shear band tip. The displacements in the figures are amplified by a deformation scale factor of 50.}
\label{lama}
\end{figure}

A mesh of the same size as that previously called \lq fine' was used to perform a simulation of a rectangular block ($H = 10$ mm, $L = 4H = 40$ mm) 
made up of a material close to shear band instability ($E_p = 300$ MPa) and 
containing a shear band (of length $H/2 = 5$ mm and constitutive thickness $2h = 0.01$ mm). 
Results are presented in Fig.~\ref{lama}. In parts (a) and (b) (the latter is a detail of part a) of this figure the overall response curve is shown of the block in terms of average shear stress $\bar{\sigma}_{12} = T/L$ ($T$ denotes the total shear reaction force at the upper edge of the block) and average shear strain $\bar{\gamma}_{12} = u_1/H$.  In part (c) of the figure contour plots of the shear deformation $\gamma_{12}$ are reported at different stages of deformation. It is clear that the deformation is highly focused along a rectilinear path emanating from the shear band tip, thus demonstrating the tendency of the shear band towards rectilinear propagation under shear loading. 

\begin{figure}[!htcb]
\centering
\includegraphics[width=130mm]{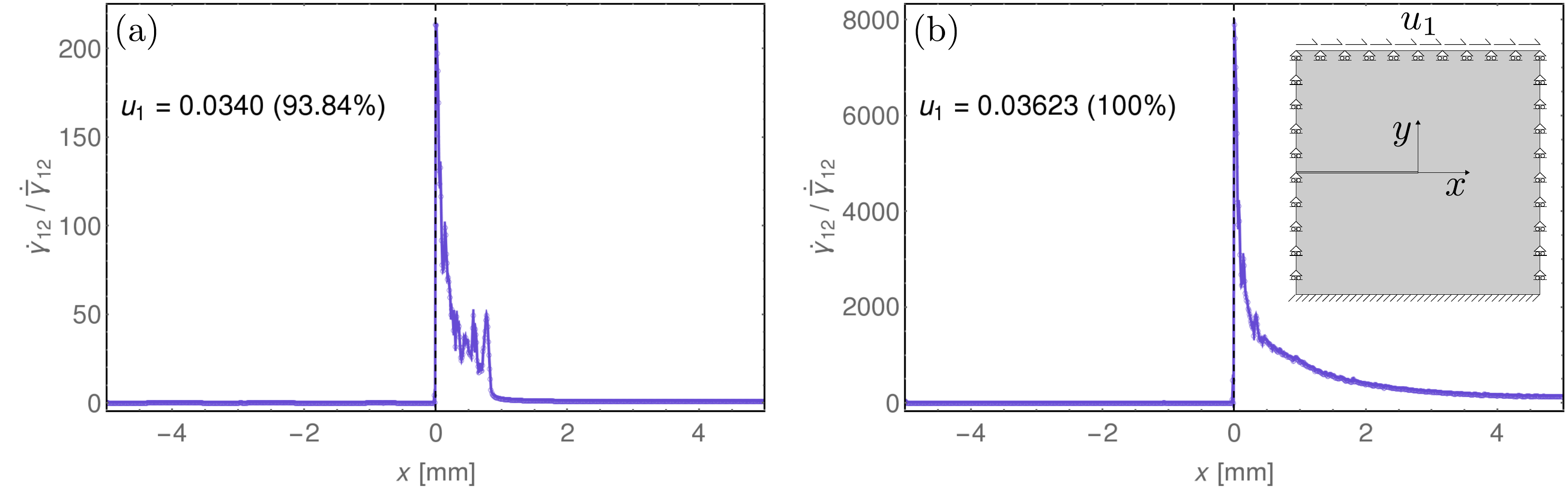}
\caption{The incremental shear strain $\dot{\gamma}_{12}$ (divided by the mean incremental shear strain $\dot{\bar{\gamma}}_{12}$) along the $x$-axis at the two stages of deformation reported in in Fig. \ref{pathcase1} and labeled there as (a) and (c). It is clear that a strong strain concentration develops at the tip of the shear band, which becomes similar to the square-root singularity that is found with the 
perturbative approach (Section \ref{sec:pert} and Fig. \ref{incrementaleps}. }
\label{peppa}
\end{figure}

Finally, the incremental shear strain (divided by the mean incremental shear strain) has been reported along the $x$-axis in Fig. \ref{peppa}, at the two stages of deformation considered in Fig. \ref{pathcase1} and referred there as (a) and (c).
These results, which have been obtained with the fine mesh, show that a strong incremental strain concentration develops at the shear band tip and becomes qualitatively similar to the square-root singularity 
found in the perturbative approach. 


\section{The perturbative vs the imperfection approach}
\label{sec:pert}

With the perturbative approach, a perturbing agent acts at a certain
stage of uniform strain of an infinite body, while the material
is subject to a uniform prestress. Here the perturbing agent is a pre-existing shear band, modelled as a planar slip surface, emerging at a certain stage of
a deformation path (Bigoni and Dal Corso, 2008), in contrast with the imperfection approach in which the imperfection is present from the beginning of the loading.

With reference to a  $x_1$--$x_2$ coordinate system (inclined at 45$^\circ$ with respect to
the principal prestress axes $x_I$--$x_{II}$), where the incremental stress $\dot{t}_{ij}$ and
incremental strain $\dot{\varepsilon}_{ij}$ are defined ($i,j=$1,2), the
incremental orthotropic response
under plane strain conditions ($\dot{\varepsilon}_{i3}=0$) for incompressible materials
($\dot{\varepsilon}_{11}+\dot{\varepsilon}_{22}=0$) can be expressed
through the following constitutive equations (Bigoni, 2012)\footnote{Note that the notation used here differs from that adopted in (Bigoni and Dal Corso, 2008), where the principal axes are denoted by $x_1$ and $x_2$
and the system inclined at 45$^\circ$ is denoted by $\hat{x}_1$ and $\hat{x}_2$.}
\begin{equation}
\dot{t}_{11}= 2\mu \dot{\varepsilon}_{11} + \dot{p}, \quad
\dot{t}_{22}= - 2 \mu \dot{\varepsilon}_{11} + \dot{p}, \quad
\dot{t}_{12}= \mu_* \dot{\gamma}_{12},
\end{equation}
where $\dot{p}$ is the incremental in-plane mean stress, while $\mu$ and
$\mu_*$ describe the incremental shear stiffness, respectively, parallel and
inclined at 45$^\circ$ with respect to prestress axes.

The assumption of a specific constitutive model leads to the definition of
the incremental stiffness moduli $\mu$ and $\mu_*$.
With reference to the J$_2$--deformation theory of plasticity (Bigoni and Dal Corso, 2008), particularly
suited to model the plastic branch of the constitutive response of ductile metals, 
the in-plane deviatoric stress can be written as
\begin{equation}
\label{dev}
t_I-t_{II}=k \varepsilon_I |\varepsilon_I|^{(N-1)}.
\end{equation}
In equation (\ref{dev}) $k$ represents a stiffness coefficient and $N\in(0,1]$ is the strain hardening exponent, describing perfect plasticity (null hardening)
in the limit $N \rightarrow 0$ and linear elasticity in the limit $N\rightarrow 1$.
For the J$_2$--deformation theory, the relation between the two incremental shear stiffness moduli can be obtained as
\begin{equation}
\mu_*=N \mu,
\end{equation}
so that a very compliant response under shear ($\mu_*\ll\mu$) is described in the limit of perfect plasticity $N \rightarrow 0$.

\begin{figure}[!htcb]
\centering
\includegraphics[width=100mm]{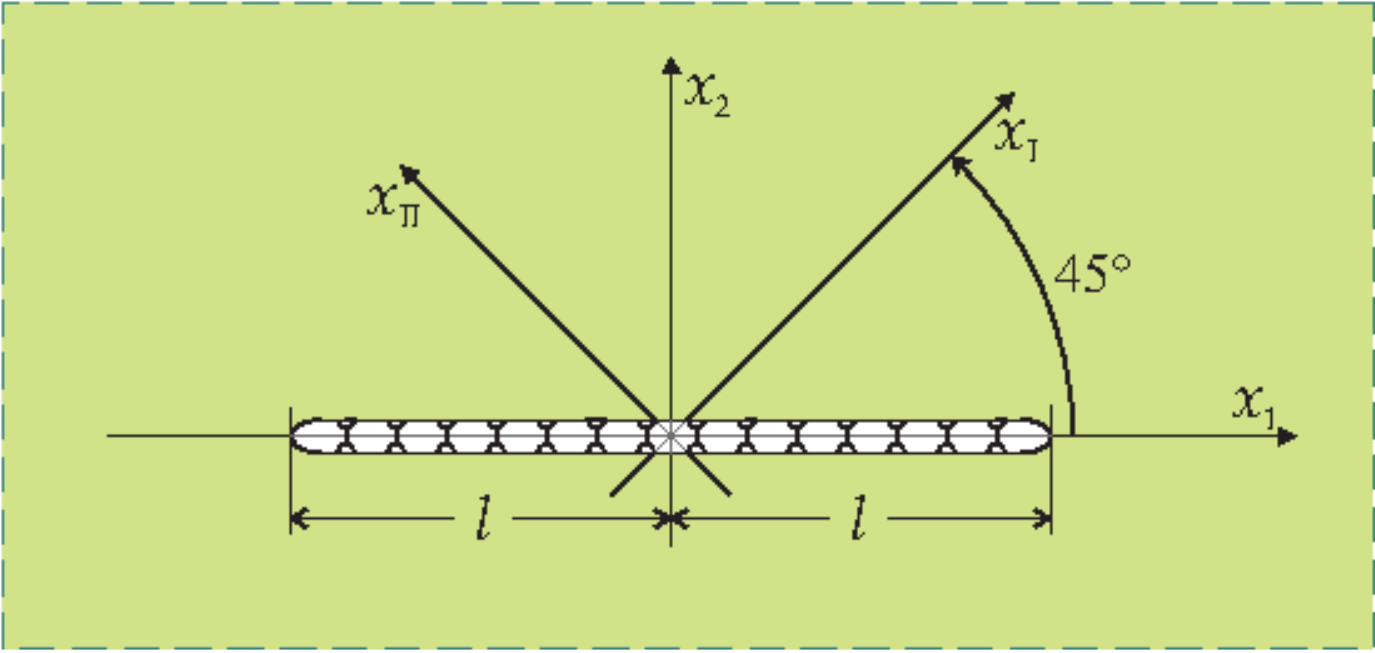}
\caption{A perturbative approach to shear band growth: a pre-existing shear band, modelled as a planar slip surface, acts at a certain stage of uniform deformation of
an infinite body obeying the J$_2$--deformation theory of plasticity}
\label{inclinatone}
\end{figure}

The perturbative approach (Bigoni and Dal Corso, 2008) can now be exploited to investigate the growth of a shear band within a solid.
To this purpose, an incremental boundary value problem is formulated for an infinite solid, containing a zero-thickness
pre-existing shear band of finite length $2 l$ parallel to
the $x_1$ axis (see Fig.~\ref{inclinatone}) and loaded at infinity through a uniform shear deformation $\dot{\gamma}_{12}^\infty$.

The incremental boundary conditions
introduced by the presence of a pre-existing shear band can be described by the following equations:
\begin{equation}
\dot{t}_{21}(x_1,0^\pm)=0, \quad
\jump{\dot{t}_{22}(x_1,0)}= 0, \quad
\jump{\dot{u}_{2}(x_1,0)} = 0, \quad
\forall |x_1|< l . 
\end{equation}
A stream function $\psi(x_1,x_2)$ is now introduced, 
automatically satisfying the incompressibility condition and defining the incremental displacements $\dot{u}_j$ as $\dot{u}_1
= \psi_{,2},$ and $\dot{u}_2=- \psi_{,1}$. 
The incremental
boundary value problem is therefore solved as the sum of 
$\psi^\circ(x_1,x_2)$, solution of the incremental homogeneous problem, and $\psi^p(x_1,x_2)$, solution of the incremental perturbed problem.

\begin{figure}[!htcb]
\centering
\includegraphics[width=160mm]{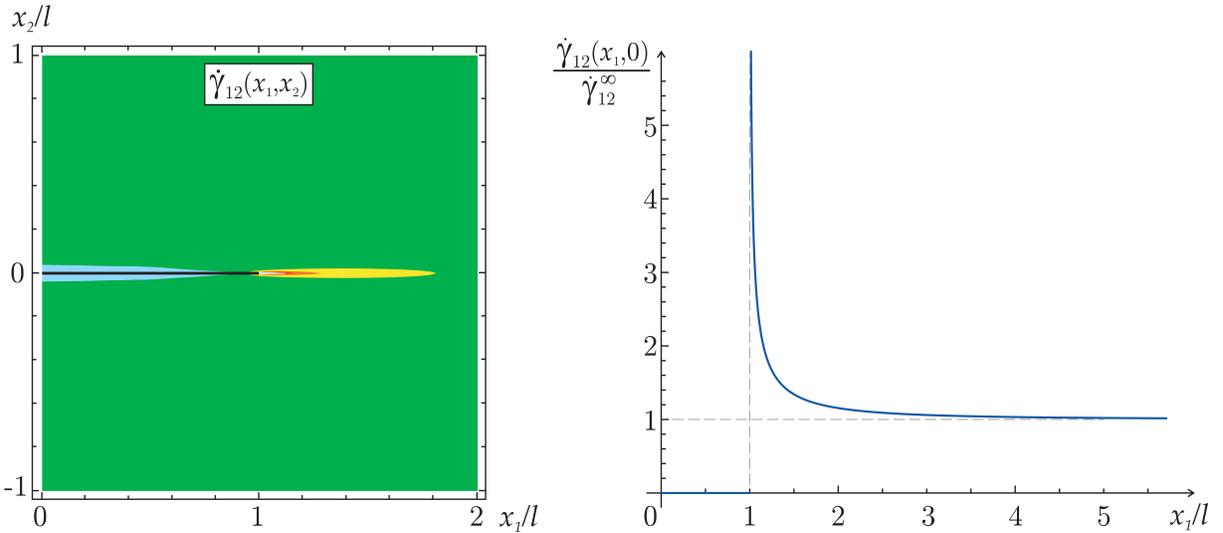}
\caption{Incremental shear strain near a shear band obtained through the perturbative approach: level sets (left) and behaviour along the $x_1$-axis (right).}
\label{incrementaleps}
\end{figure}

The incremental solution is reported in Fig. \ref{incrementaleps} for a low hardening exponent, $N=0.01$, as a contour plot (left) and as a graph
(along the $x_1$-axis, right) of the incremental shear deformation $\dot{\gamma}_{12}$
(divided by the applied remote shear $\dot{\gamma}_{12}^\infty$).
Note that, similarly to the crack tip fields in fracture mechanics,
the incremental stress and deformation display square root singularities
at the tips of the pre-existing shear band.
Evaluation of the solution obtained from the perturbative approach analytically confirms the conclusions drawn from
the imperfection approach (see the numerical simulations reported in Fig. \ref{contour_e12_case1} and \ref{lama}), in particular:
\begin{itemize}

\item It can be noted from Fig.~\ref{incrementaleps} (left) that the incremental deformation
is highly focussed along the $x_1$ direction, confirming that the shear band grows rectilinearly;

\item The blow-up of the incremental deformation observed in the numerical simulations near the shear band tip (Fig. \ref{peppa}) is substantiated by the theoretical
square-root singularity found in the incremental solution (Fig.~\ref{incrementaleps}, right).

\end{itemize}

\section{Conclusions}

Two models of shear band have been described, one in which the shear band is an imperfection embedded in a material and another in which the shear band is a perturbation which emerges during a homogeneous deformation of an infinite material. These two models explain how shear bands tend towards a rectilinear propagation under continuous shear loading, a feature not observed for fracture trajectories in brittle materials. This behaviour is a basic micromechanism of failure for ductile materials.
These models show also a strong stress concentration at the shear band tip, which can strongly promote shear band growth.

\section*{Acknowledgments}
D.B., N.B. and F.D.C. gratefully acknowledge financial support from the ERC Advanced Grant \lq Instabilities and nonlocal multiscale modelling of materials'
FP7-PEOPLE-IDEAS-ERC-2013-AdG (2014-2019). A.P. thanks financial support from the FP7-PEOPLE-2013-CIG grant PCIG13-GA-2013-618375-MeMic.

\end{document}

%% file: definitions.tex
\newcommand{\beq}{\begin{equation}}
\newcommand{\eeq}{\end{equation}}

\newcommand{\beqar}{\begin{eqnarray}}
\newcommand{\eeqar}{\end{eqnarray}}
\newcommand{\bit}{\begin{itemize}}
\newcommand{\eit}{\end{itemize}}
\newcommand{\benum}{\begin{enumerate}}
\newcommand{\eenum}{\end{enumerate}}
\newcommand{\barr}{\begin{array}}
\newcommand{\earr}{\end{array}}
\def\ds{\displaystyle}

\def\XXint#1#2#3{{\setbox0=\hbox{$#1{#2#3}{\int}$}
   \vcenter{\hbox{$#2#3$}}\kern-.5\wd0}}

\def\b0{\mbox{\boldmath $0$}}

\def\bt{\mbox{\boldmath $t$}}
\def\bu{\mbox{\boldmath $u$}}

\def\bK{\mbox{\boldmath $K$}}

\def\f0{\ensuremath{\mathbb{O}}}

\def\IJF{{\it Int.\ J.\ Fracture}\ }

\def\IJSS{{\it Int.\ J.\ Solids Struct.}\ }

\def\JMPS{{\it J.\ Mech.\ Phys.\ Solids}\ }